\documentclass[submission]{eptcs}
\usepackage{oz}
\providecommand{\event}{Refine 2015} 
\usepackage{breakurl}             

\title{Big Data Refinement}
\author{Eerke A. Boiten
\institute{School of Computing and Cybersecurity Centre\\
University of Kent, UK}
\email{E.A.Boiten@kent.ac.uk}}
\def\titlerunning{Big Data Refinement}
\def\authorrunning{E.A. Boiten}
\begin{document}
\maketitle

\begin{abstract}
``Big data" has become a major area of research and associated funding, as well as a focus of utopian thinking. 
In the still growing research community, one of the favourite optimistic analogies for data processing is that of the oil refinery, extracting
the essence out of the raw data. Pessimists look for their imagery to the other end of the petrol cycle, and talk 
about the ``data exhausts" of our society.

Obviously, the refinement community knows how to do ``refining".
This paper explores the extent to which notions
of refinement and data in the formal methods community relate to the core concepts in ``big data". In particular, can
the data refinement paradigm can be used to explain aspects of big data processing?
\end{abstract}

\section{Introduction}
``Big data" has been a topic of great interest internationally for a few years now, and the UK government has declared it to be one
of the ``eight great technologies\footnote{\tt https://www.gov.uk/government/publications/eight-great-technologies-big-data}". As a consequence, it is opportune for researchers and institutes to consider how they can engage or, more cynically: rebrand, in order to further the research agenda and profit from available funding opportunities.

The Open Data Insititute in the UK presents\footnote{See {\tt https://github.com/theodi/data-definitions},
also for considerations on how ``big data" relates to ``open data" and ``personal data".} two definitions of big data:
\begin{quote}
\begin{enumerate}
\item[(i)] data that you cannot handle with conventional tools, or
\item[(ii)] a term used as a vague metaphor for solving problems with data.
\end{enumerate}
\end{quote}
The refinement community could engage with the big data research field along the lines of
 either of those definitions.

The first, more technical, definition is the one the UK government was mainly looking at, considering problems that require ``petabytes" of
data to be processed.  Typical descriptions of big data in this context refer to three (or more!) `V's \cite{Laney01}: volume, velocity and variety. The first two highlight that not only there will be a lot of data, but it may also be produced at a persistently high rate. The last of these refers to the possible lack of uniform structure in the data.
Research communities in networks, databases, programming languages, and precursor bandwagon areas like ``grid computing" and
``cloud computing" have already moved into this technical field of ``big data", leading to a substantive
research effort on the side of ``engineering" of big data.
From a formal methods perspective, the application of big data tools like Hadoop and MapReduce comes with its own verification requirements. Indeed, some research has already been done to explore this \cite{DBLP:conf/services/ReddyFLD0K13}, e.g.\ using CSP
\cite{DBLP:conf/tase/SuYZL09},  Coq \cite{Ono11} or QuickCheck and VDM \cite{DBLP:conf/ipps/Kusakabe11}. It is so
far less clear whether there are formal methods challenges in this area which fundamentally relate to the {\em paradigm} rather than to the verifiability of the use of big data programming tools.

In this paper, however, we concentrate on the second definition. It is arguably the one with a wider public appeal. It shares a lot of its promise, attraction, and even substance with ideas about ``artificial intelligence" that have
been part of popular science for some sixty years now. More to the
point, it is also the one that asks the most interesting questions of refinement. In this view, ``big" does not refer to the size of the data so much as to its perceived potential. 

There are two metaphors which are commonly used for big data processing - both of them inspired by the fossil fuel cycle. In what
follows, we will show that each naturally leads to a different refinement perspective. First, in Section \ref{sec:exhaust} we discuss
the {\em data exhaust} analogy, and how it relates to the idea of {\em output refinement}. Then, in Section \ref{sec:refinery} we
consider the {\em data refinery} analogy. An understanding of this in refinement terms, related to the simplified stance that
big data is ``just" statistics, requires probabilistic notions of refinement. Following this realisation, we take
a second look at
the ``data exhaust" view, in Section \ref{sec:second}. In Section \ref{sec:conc} we reflect on the relevance of this paper's attempts
to relate data refinement and big data.

\section{The data exhaust and output refinement}\label{sec:exhaust}
The data exhaust scenario is one that is causing some software producers and many commercial
and governmental organisations serious anxiety. The line of reasoning goes:
our systems (or our apps) are generating so much data -- location data, usage logs, maybe all the way to full-fledged digital surveillance --
but we are not extracting any information from this. Such information would allow us to increase our knowledge and
subsequently improve our processes or effectiveness, so this must be a problem.

This was phrased explicitly using the traditional\footnote{So traditional that there does not seem to be a definitive reference.} distinction in information science between ``data", ``information", and ``knowledge", where information adds an interpretative meaning to data, and knowledge is about productive use of information. The data ``exhaust"
acknowledges the gap between big data and the elusive ``big information". Data is being produced as a side
effect of the core process, but its information content is not uncovered let alone exploited.

A complete modelling of the exploitation of the data exhaust would be in two steps: first, information would need to be extracted from the data, and then likely this would need to be fed back into the system somehow in order to improve its performance. The first step is essentially a notion of {\em output refinement}.

Using the standard Z schema representation of operations, the simplest 
definition of output refinement is as follows. (This is a simplification of \cite[Def 10.10]{DB14}, IO-downward simulation. It considers a single operation with no inputs before nor after refinement (a trivial input transformer)
and the identity retrieve relation on the state.)
\newtheorem{definition}{Definition}\label{def:outref}
\begin{definition}[Plain Output Refinement] The operation $AOp$ on state $State$  is output refined by $COp$ operating on the same state if an IO transformer\footnote{See \cite{DB14} for full detail, in summary: the signature of a schema $S$ is defined as $S \vee \neg S$, effectively replacing any predicates of $S$ by $\true$. The input signature
$?S$ and output signature $!S$ are restrictions of the signature of $S$ to inputs and outputs, respectively. Any schema $S$ with only
inputs and outputs, i.e.\ one such that
$\Sigma S = ?S \wedge !S$, is an IO-transformer. Its converse $\overline{S}$ swaps all inputs and outputs, so e.g.\ $a!$ in $S$ is consistently substituted by $a?$ in $\overline{S}$, and more generally $? \overline{S}
= !S$. A schema $T$ is an output transformer for $S$ if it its inputs match exactly the outputs of $S$, i.e.\ $!S =\overline{?T}$. In that case, post-composition using the standard $\zpipe$ operator has the intuitive meaning that
it hides $S$'s original output as well as the inputs of $T$ that they are identified with.} $OT$ exists such that:
\begin{itemize}
\item $OT$ is a total injective output transformer for $AOp$;
\item $\forall State \dot \pre AOp \Rightarrow \pre COp$
\item $\forall State; State'; !COp \dot \pre AOp \wedge COp \Rightarrow AOp \zpipe OT$
\end{itemize}
\end{definition}
The semantic justification for this (see the derivation of IO-refinement in \cite{DB14}) is from the standard
assumption for Z refinement, namely that inputs and outputs
are visible to the outside world, as well as the operations that are executed -- but the internal state is not
observable directly. 

Due to this visibility of
outputs in the semantics, implicitly in the context there has to be an ``original output transformer" which records how the output actually produced in a specification that has undergone
output refinement can be transformed into the output (type) stated in the original specification. This transformer needs
to be functional from modified to original output: for any modified output value, we need to establish unambiguously which
original output value it represents. (Consider for example a concrete output of a display screen made up of
pixels, and an abstract output type of a single digit.) In \cite{DB14} we also called this the ``every sperm is sacred" principle to ensure we had
a Monty Python reference in an academic textbook. The injectivity requirement of the output transformer in Definition \ref{def:outref} follows from this, see the derivation of IO refinement in \cite{DB14}.

So does this definition help us to validate use of the data exhaust? Well, {\em creating} a data exhaust is captured by output refinement. The transformation which adds extra outputs satisfies the
criteria of Definition \ref{def:outref}, using a trivial output transformer which copies all existing outputs
while not constraining any additional ones. The intuition that it corresponds with is: implementing a component with another that provides extra output wires which we then just do not connect to anything. 

On the other hand, if the data exhaust is already present as an output, replacing it with some useful value derived from it is not normally an injective operation, and so output refinement does not hold in general. Thus, we have to conclude that ``big data" processing in the ``data exhaust" perspective is {\em not} an instance of refinement. 

In fact, output refinement likely holds in the opposite direction. Going from the extracted information to the data
exhaust is likely to be injective (i.e., its converse is functional). Thus, the data exhaust view of big data 
represents {\em output abstraction} (in the semantic sense -- rather than the syntactic sense \cite{BoiIOabs}). This makes perfect sense, but does not do full justice to the ``exhaust" analogy. More on that later.

\section{The data refinery and probabilistic refinement}\label{sec:refinery}
The data refinery metaphor is of a much more optimistic nature. In this case, the narrative is that when
we have or collect so much ``raw" data, it will certainly contain the answers to all questions we might want to ask.

\strut\\
{\em Political interlude.}
This has significant
political consequences \cite{Pasquale15}, in areas like communications surveillance, targeted marketing \cite{Borgesius15}, and medical and genomic
data.  In particular, collecting the data is immediately justified by the promise of all the questions it will answer, and big data, even personal data, becomes
a resource in and by itself. If only we collect enough communications data, we will be able to identify all terrorists and
thwart their evil plans. Similarly, any sizable health database will have embedded in it the cure for cancer, just waiting for the clever
big data techniques to extract it. Google's Larry Page has been making this kind of case for health data \cite{Boiten14}. {\em End political interlude.}

\strut\\
If we view this as a refinement scenario, the ``raw data" is the {\em state} of the system. This means that we need not worry,
as we did in the data exhaust scenario, about throwing any of it away, as none of it is visible to the outside world to start with.
So what does the data refinery achieve? It starts out in a world where we have the data but not the answer to the question. To simplify
the picture, let us assume that the question is a binary one. Thus, our ``abstract" specification is
\begin{schema}{RawIgnorance}
b, b': BigData \\
a!: Answer \ST
b'=b \\
a! = yes \vee a!= no
\end{schema}
Modelling ignorance as non-determinism may not be sophisticated, but at least it is simple!

The transformed specification, making use of clever big data processing techniques, might look as follows. Let's assume that $StructuredData \subseteq BigData$\footnote{Thanks to Mark Utting for suggesting to reflect the clever processing in a change of type of state as well as a more precise output. The technical constraint that $StructuredData$ is included in $BigData$  is justifiable in the interpretation, and avoids complex arguments about retrieve relations and multiple
executions of the $MachineLearn$ operation.}.
\begin{schema}{MachineLearn}
b: BigData \\
b': StructuredData \\
a!: Answer \ST
b'= cleverprocessing(b) \\
a! = answer(b')
\end{schema}
Assuming the function $answer$ returns $yes$ or $no$, this transition is a trivial data refinement, given that $RawIgnorance$ is the weakest possible specification for a yes-or-no answer, and ignores the state. But it does not do big data processing much justice, and as a corollary this modelling process cannot be very enlightening. 

A frequently expressed view is that big data is ``just" the application of statistics -- and indeed many machine learning methods produce outputs with a degree of confidence, a statistical indication of how ``right" the answer is
felt to be. To model this,  we could assume that the function $answer$ 
returns a probability distribution over the values ``yes" and ``no". This would represent a probabilistic refinement, e.g.\ in the framework developed
by McIver and Morgan \cite{McIverMorgan04}, as non-determinic choice is refined by probabilistic choice. Our modelling of ignorance
becomes less naive in that context, as non-determinism (``possibilism") fairly describes all possible probabilistic outcomes.

However, there is still a mismatch that way. For example, if our confidence in the conclusion that the answer is $yes$ is 93\%, what probability
distribution describes that best?A first cut is this:
\[ yes \ _{0.93} \oplus_{0.07}\, no \]
which returns $yes$ with probability $0.93$ and $no$ with probability $0.07$. It is {\em not} the right
answer -- because it suggests a certainty over the remaining 7\% that we do not actually possess. 

The mixture of probability and
non-determinism in the McIver and Morgan framework would allow us to specify a non-deterministic choice
(represented by $\sqcap$) of $yes$ and $no$ for the
remaining 7\%:
\[ yes \ _{0.93} \oplus_{0.07} (yes \sqcap no) \]
Maybe that is a correct encoding of a belief in $yes$ with 93\% confidence, but as a specification it is getting
clumsy.

A cleaner solution to this, which associates the confidence with the {\em refinement judgement} itself rather than with the outcome of the judgement, is presented in Mingsheng Ying's probabilistic logic for probabilistic programs \cite{Ying03}. This does not only associate probabilities
with program behaviours, but also with judgements of correctness and refinement. In such a framework, statistical machine learning
is modelled as a refinement away from ignorance -- with any degree of confidence attached as an attribute of the refinement step itself. So rather than refining to a probabilistic specification, we would e.g.\ have
\begin{schema}{Result}
b: BigData \\
b': StructuredData \\
a!: Answer \ST
b'= cleverprocessing(b) \\
a! = yes
\end{schema}
presented as a ``93\% refinement" of $RawIgnorance$.

So, good news: big data processing has been successfully modelled as data refinement. Except, of course, that normally when
we talk of data refinement we consider different data representations, related by some retrieve relation or coupling invariant.
In this model of ``big data", that played no role to speak of. How come? 
When we talk of ``data types" and ``data refinement" in formal methods, it seems we don't actually mean ``data" in the sense of the data-information-knowledge hierarchy. The role of data is as a way of representing information, formal methods ``data" is already
``useful". (In algebraic data type theory, we even explicitly talk about ``junk", values in the model that we would like to
keep outside our considerations.) 
To fall in line with the common usage of ``data" and ``information", maybe we should in the future be talking of ``information refinement" or ``information representation refinement" instead of ``data refinement".

\section{The data exhaust, revisited}\label{sec:second}
The perspective that data-as-in-big is not the same as data-as-in-refinement, in combination with probabilistic refinement ideas,
invites a return to the data exhaust analogy. In particular, another way to look at the failing injectivity criterion for output refinement
is that it stops us
from throwing away information. (Or from reducing entropy, if you like.) But actually, reducing some of the data exhaust should then not be forbidden: the data exhaust {\em was never information in the first place}.

So how can we take a more sophisticated view of this? The data exhaust could be viewed as data that hides a little bit of information
in a cloud of noise. Looking at machine learning in reverse, the exhaust forms a statistical obfuscation of the relevant information.
In the world of information (refinement), we're not interested in the exact value after obfuscation -- different obfuscations of the
same data should be considered equal. This leads into all kinds of interesting directions.

We could imagine {\em noisy refinement}, where output values are constructed from types $SIGNAL$ and $NOISE$
using an operation
\begin{axdef}
out: SIGNAL \cross NOISE \rightarrow SIGNAL
\ST
\forall x,y: NOISE; a:SIGNAL \dot
\exists z:NOISE \dot out(out(a,x),y) = out(a,z)
\end{axdef}
(adding noise is idempotent), where observations are of type $SIGNAL$ with an original output transformer
(see the discussion on IO refinement in Section \ref{sec:exhaust})
\begin{schema}{OOT}
ot?, oo!: SIGNAL
\ST
\exists x:NOISE \dot ot? = out(oo!,x)
\end{schema}
i.e., the concrete output is the original output with added noise. All refinement would then be ``modulo addition of noise to outputs".

An interesting question would be whether
this then could be related to ``noisy channels" in theoretical and quantum cryptography
\cite{Crepeau97}, as these allow cryptographic constructions that are impossible on perfect channels.

\strut\\
A final wild thought would be to look not just at yes-no questions, but at discovering relations in the big data.
Machine learning and regression analysis to determine which relations hold between values of different variables
could be seen as ``reverse engineering", or: if these are the values of inputs and outputs, what is the abstract
data type that produces these?
If this also happens with degrees of confidence, we could either use approximate refinement relations \cite{BD05} in
contexts where we know how to arbitrarily increase precision, or otherwise quantified refinement \`a la Mingsheng
Ying as mentioned above.

\section{Conclusions}\label{sec:conc}

This paper was trying to understand a large, popular, and vaguely defined bandwagon topic from the perspective of a very 
well-defined and well-established but narrow research niche. A differing lack in precision of terminology was always likely
to trip that effort up. Now, is there such a topic as ``big data refinement"? Probably not, as the two areas really mean different things when they use the word ``data".  However, looking at the two areas side by side was useful in clarifying where the difference between ``data" and ``information" really lies in both areas.

Finally, an interesting conclusion from this analysis is that, with the right probabilistic generalisations in place,
refinement can be seen to subsume machine learning: going from a situation where we have raw data and
an unanswered question to one where we have achieved an answer with a degree of confidence.

\bibliographystyle{eptcs}
\bibliography{bigdatarefine}
\end{document}